\documentclass{aa}  
\usepackage{float}
\usepackage{pgfplots}
\pgfplotsset{compat=1.15}
\usepackage{graphicx}

\usepackage{txfonts}
\usepackage{makecell}
\usepackage{tikz}
\usepackage{tikz-3dplot}

\usepackage{amsmath}
\usepackage{xcolor}
\usepackage[usenames,dvipsnames]{xcolor}
\definecolor{MyDeepBlue}{HTML}{08519C}
\usetikzlibrary {arrows.meta}

\usepackage[colorlinks=true,
            linkcolor=blue,    
            citecolor=blue,      
            urlcolor=blue,        
            pdfborder={0 0 0}     
           ]{hyperref}
\begin{document}

   \title{The evolution of the spin alignments of dark matter halos in the cosmic web}

   \subtitle{}

 \author{
  David Tobar\inst{1}
  \and Rory Smith\inst{1,2}
  \and Antonio D. Montero-Dorta\inst{3}
  \and Katarina Kraljic\inst{4}
  \and Pablo López\inst{5,6}
}

\institute{
   Departamento de Física, Universidad Técnica Federico Santa María, Av. España 1680, Valparaíso, Chile\\
  \email{david.tobar@usm.cl}
  \and
  Millenium Nucleus for Galaxies (MINGAL)
  \and
   Departamento de Física, Universidad Técnica Federico Santa María, Vicuña Mackenna 3939, San Joaquín, Santiago de Chile, Chile
  \and
   Observatoire Astronomique de Strasbourg, Université de Strasbourg, CNRS, UMR 7550, F-67000 Strasbourg, France
   \and
   Observatorio Astronómico de Córdoba, Universidad Nacional de Córdoba (UNC), Francisco N. Laprida 854, Córdoba, Argentina
   \and
   Instituto de Astronomía Teórica y Experimental, CONICET-UNC, Laprida 922, Córdoba, Argentina\\
}

   \date{}

  \abstract
   {}
   {We investigate the evolution of dark matter halo spin alignments with respect to cosmic filaments and explore how halo mass, proximity to filaments, and major mergers affect their orientation over time.}
   {We performed a suite of dark matter-only zoom-in N-body simulations centered on ten filaments extracted from a cosmological box using the 1DREAM structure finder. The zoom-in technique allowed us to resolve low-mass dark matter halos within each filament while preserving the large-scale environment. Halos were identified with the AMIGA Halo Finder, and their evolutionary histories were reconstructed to trace the spin, shape, and distance to the filament from redshift $z = 1$ to $z=0$.}
   {Despite limited statistics, we confirm a strong mass-dependent alignment signal: low-mass halos tend to align parallel to the filament, while high-mass halos favor perpendicular orientations. The perpendicular alignments become dominant toward the highest halo masses in our sample, around $\log_{10}(M_{\mathrm{h}}/h^{-1}\mathrm{M_\odot}) \sim 12$. In addition, we identify major mergers as events capable of producing sharp spin reorientations and temporary transitions toward more prolate halo shapes, particularly for massive halos located near the filament core; this suggests a preferential merger direction within filaments.}
   {Overall, halo mass emerges as the primary factor governing spin--filament alignments in our sample. By analyzing the global evolution, we find that the average orientations measured at $z=0$ do not differ significantly from those at $z = 1$, indicating that the bulk of the present-day spin configuration was largely established at earlier stages of halo evolution. Major mergers, although relatively rare, represent one of the few mechanisms capable of disrupting this initial alignment.}

\titlerunning{Dark matter halo alignments}
\authorrunning{David Tobar et al.}
   \maketitle

\section{Introduction}
 \label{Introduction}

The current paradigm of cosmic structure formation establishes that the large-scale structure (LSS) of the Universe arises from small primordial density fluctuations. In the linear regime, these perturbations grow according to linear perturbation theory \citep[e.g.,][]{1980lssu.book.....P}. As the density contrast increases, nonlinear gravitational evolution leads to the formation of the cosmic web \citep{1996Natur.380..603B}, which is composed of voids, sheets, filaments, and nodes. Within this hierarchical framework, matter flows toward denser environments, where gravitational collapse forms virialized dark matter (DM) halos, the fundamental building blocks of galaxies and clusters \citep{1980lssu.book.....P}.

According to the tidal torque theory (TTT), the angular momentum of proto-halos is generated during the early stages of structure formation by tidal torques exerted by the surrounding mass distribution arising from the misalignment between the inertia tensor and the local tidal field \citep{Hoyle1951,Peebles1969,Doroshkevich1970,White1984}. These tidal shears, produced by asymmetries in the gravitational potential, induce coherent torques that efficiently build up angular momentum in the linear regime. However, tidal torques become progressively inefficient once a proto-halo decouples from the cosmic expansion and undergoes collapse. After this stage, the proto-halo separates from neighboring perturbations and its moment of inertia decreases rapidly, effectively freezing the angular momentum acquired during the TTT phase \citep{2002MNRAS.332..325P,2002MNRAS.332..339P}.

In this context, the alignment of DM halo spins with the surrounding LSS has been extensively investigated using cosmological N-body simulations. Early studies aimed to characterize these alignments \citep[e.g.,][]{Lee_2000,Hatton&Ninin,Faltenbacher:2002tk,10.1111/j.1365-2966.2006.11318.x,10.1111/j.1365-2966.2008.14244.x,Zhang_2009}. Currently, there is a general agreement on the alignment of DM halos, i.e., that the orientation of their spins is not random but correlated with both the LSS in which they reside and their mass. Regarding the orientation with respect to filaments -- the main focus of this work -- more recent analyses have shown that the orientation of the halo spin vector is mass-dependent \citep[e.g.,][]{10.1093/mnras/stv1570,Wang_2018,10.1093/mnras/sty2270,2018MNRAS.481.4753C,2020MNRAS.493..362K,10.1093/mnras/stab411}. 
This mass-dependent alignment behavior, often related to a phenomenon known as spin-flip, suggests a transition in the angular momentum acquisition process that depends on the halo's mass and environment \citep{Codis_2012,Trowland_2013,2013MNRAS.428.2489L,Pichon_Codis_Pogosyan_Dubois_Desjacques_Devriendt_2014}.\ This leads to the so-called transition mass, where the alignment behavior of halos changes. Theoretically, this mass-dependent inversion arises from the interplay between large-scale tidal flows and local accretion geometry \citep{2013MNRAS.428.2489L,2015MNRAS.446.2744L}. Low-mass halos acquire parallel spin through the winding of flows around the filament \citep{Codis_2012}, whereas massive halos grow primarily via mergers and diffuse accretion along the filament spine, generating perpendicular angular momentum \citep{10.1093/mnras/stu1216}. This directional accretion, which also drives the anisotropic and planar alignments of satellite galaxies \citep{10.1093/mnras/stv919, 10.1093/mnras/stv1302}, is expected to generate angular momentum perpendicular to the filament axis, leading to the observed mass-dependent transition \citep[e.g., ][]{10.1111/j.1365-2966.2011.20275.x,10.1093/mnrasl/slu106,2015MNRAS.446.2744L,10.1093/mnras/sty2270,Krolewski_2019,10.1093/mnras/stab411}.

Moreover, it has been demonstrated that the transition mass is not universal but depends on the environmental scale, including the filament's thickness and density.  \citet{2014MNRAS.440L..46A} showed that the transition mass varies with the hierarchical level of the filament. Similarly, \citet{10.1093/mnras/sty2270,10.1093/mnras/stab411} presented evidence that the transition mass increases with filament diameter, a behavior linked to the accretion dynamics described by \citet{2018PhDT.......278B}.

A closely related framework is that of intrinsic alignments (IAs), which describe correlations between galaxy and halo orientations induced by large-scale tidal fields \citep{hirata_2004,troxel_2015,10.1093/mnras/stx2846}. IAs are both a probe of structure formation and a major systematic in weak-lensing analyses \citep{hikage_2019,fabbian_2019,chisari_2025_review}. In this context, studying the evolution of halo spin–filament alignments, and their connection to mergers and anisotropic accretion, provides insight into the physical processes underlying IAs.

For shape alignments, another relevant aspect of the problem, the interpretation is relatively straightforward because halo shapes are largely imprinted by the anisotropic ellipsoidal collapse driven by the large-scale tidal field \citep[e.g.,][]{10.1111/j.1365-2966.2005.09934.x,2011MNRAS.414.2029P,2014MNRAS.443.1090F}. These processes naturally induce a preferred orientation of halos with respect to the surrounding large-scale tidal field. In particular, the minor axis of halos tends to be preferentially perpendicular to the filament \citep[][]{2001ApJ...555..106L,aragon-calvo,10.1111/j.1365-2966.2006.11318.x,10.1093/mnras/stw1247,10.1093/mnras/sty2270,10.1093/mnras/stab451}. Similar trends have also been reported for galaxies and galaxy clusters \citep{2012MNRAS.423..856S,2018MNRAS.481.4753C}. This is a consequence of the sequential nature of collapse: first, a 1D collapse forms sheets, then a 2D collapse forms filaments, and finally a 3D collapse occurs as matter flows along filaments into nodes \citep[e.g.,][]{1zeldovich1970, 10.1093/mnras/stv1570}.

Observational studies have shown that galaxy spin orientations are not randomly distributed with respect to the surrounding filamentary network \citep{lee2002detectiongalaxyspinalignments,Lee_2007}. Instead, the spin--filament alignment signal depends on both the galaxy's mass and morphology. Massive early-type red galaxies tend to have spins preferentially perpendicular to their host filaments, whereas lower-mass late-type systems more often exhibit a parallel alignment or even antiparallel alignments, and some populations show no significant alignment at all \citep{2008MNRAS.389.1127P,10.1111/j.1365-2966.2010.17202.x,10.1093/mnras/sts162,Tempel_2013,Zhang_2013,Zhang_2015,10.1093/mnras/stv2930,refId0,2020MNRAS.493..362K,KraljicKatarina2021SM3s,Rong_2025,Wang_2025}. Furthermore, recent theoretical studies suggest that the local environment plays a highly active role in shaping these orientations. For instance, \citet{storck} used a novel ``splicing'' technique to show that the spin orientation of Milky Way–mass halos is highly sensitive to environmental coupling, exhibiting fluctuations of up to $80\%$ depending on their distance to a filament.

The main goal of this study was to investigate the physical drivers of halo spin alignment and to determine whether the signal observed at $z=0$ reflects intrinsic evolution or external mechanisms, such as the influence of filaments \citep{storck} or major mergers (MMs). To this end, we analyzed the dynamical evolution of halos within ten different filaments using a suite of DM-only zoom-in simulations, which provide high mass resolution while preserving the large-scale environment, allowing us to resolve halos in the mass range $10^{9.5}$–$10^{13.5}~h^{-1}\mathrm{M_\odot}$. Our work benefits from the fact that we analyzed a restricted sample of relatively straight filaments with similar geometrical characteristics; this allowed us to distinguish the effects of halo mass and dynamical evolution from variations in filament properties.

Although this study focuses exclusively on DM halos, these structures define the gravitational potential wells in which galaxies reside. In particular, we explored how MMs influence halo orientations, and whether certain halos maintain their alignment over time while others undergo significant reorientation. We examined the spatial distribution of halos that preserve or lose their alignment, and investigated if proximity to filaments tends to enhance alignment.

The article is structured as follows. Section \ref{methods} describes the procedure used to build our suite of zoom-in simulations, including halo identification and evolution. Section \ref{results} presents our main results. First, we analyze alignments at redshift $z=0$. Then, we consider different subsets of halos to study the drivers of spin alignments that could affect the evolution of DM halos in their cosmic filament environment. Section \ref{discussion} explores the physical implications of our findings and discusses alternative interpretations of the results. Finally, Sect. \ref{conclusion} summarizes our conclusions and outlines the physical hypotheses and future research avenues opened by this study.

\section{Setup}
\label{methods}
\subsection{N-body zoom-in simulation}
\label{zoomin}
We performed an initial cosmological N-body simulation consisting of $128^3$ equal-mass DM particles within a cubic box of side length $100~h^{-1}\mathrm{Mpc}$ (hereafter referred to as the level-7 simulation, since it considers $2^7 = 128$ particles in each side of the box). Each DM particle has a mass of $m_{\mathrm{DM}} \approx 5.21 \times 10^{10}~h^{-1}\mathrm{M_\odot}$.
The initial conditions were generated using the MUlti-Scale Initial Conditions (MUSIC) code \citep{Hahn_2011}, adopting the standard $\Lambda$ cold dark matter (CDM) cosmological model with parameters $\Omega_m = 0.315$, $\Omega_b = 0.049$, $\Omega_\Lambda = 0.685$, $H_0 = 67.4~\mathrm{km~s^{-1}~Mpc^{-1}}$, and $\sigma_8 = 0.811$, consistent with the Planck 2018 base-$\Lambda$CDM results \citep{planck}, initialized at redshift $z=40$. The simulations were evolved down to $z=0$ using the publicly available adaptive mesh refinement code RAMSES \citep{2002A&A...385..337T}.

To identify filamentary structures in the simulation volume, we applied the \textsc{1DREAM} framework \citep{CANDUCCI2022100658} to the $z=0$ snapshot of the level-7 simulation. 1DREAM is a novel machine-learning toolbox designed to robustly recover low-dimensional manifolds (such as filaments) in noisy environments \citep[e.g.,][]{10.1093/mnras/stad428,2024A&A...690A..92R}. The structure extraction pipeline consists of four sequential steps:
\begin{itemize}
\renewcommand\labelitemi{$\bullet$}

    \item LAAT (Locally Aligned Ant Technique): A swarm-intelligence algorithm that uses simulated pheromones to highlight high-density regions and remove background noise particles that do not belong to any structure.
    \item MBMS (Manifold Blurring Mean Shift): A denoising algorithm that iteratively moves the remaining particles toward the local density ridge, effectively collapsing the structure onto its central axis.
    \item DimIndex (Dimensionality Index): A classifier that computes the local dimensionality of the manifold, allowing us to distinguish between clusters and filaments.
    \item Crawling: A graph-based algorithm that traces the connected 1D skeleton (spine) of the filament by linking the density peaks identified in the previous steps.
\end{itemize}

In this initial step, we employed only the LAAT module to remove noisy particles and enhance the contrast of LSSs. From the resulting filament network, we selected a sample of 10 filaments through visual inspection, requiring (i) their curvature to be negligible, (ii) their lengths to be comparable, where the filament length is defined as the distance between the two ends of the filament spine, and (iii) both ends of each filament to be connected to group-mass halos with $12.8 \leq \log_{10}(M_\mathrm{h}/h^{-1}\mathrm{M_\odot}) \leq 13.7$ (see Sect.~\ref{halodata} for the halo identification process). The selected filament sample is listed in Table~\ref{filament info}, where we report some of its main properties, such as the filament length, the number of DM particles enclosed within each filament in the level-7 simulation, and its particle number density, computed within cylindrical volumes of radius $3~h^{-1}\mathrm{Mpc}$ around the filament spine. As can be seen, the selected filaments have similar properties.

\begin{table}[h!]
\centering
\caption{Properties of the selected filaments at redshift $z=0$ in the level-7 simulation, including the filament length (in $h^{-1}\mathrm{Mpc}$), the number of particles associated with each filament, and the particle density computed assuming a cylindrical volume with a radius of $3~h^{-1}\mathrm{Mpc}$.}
\label{filament info}
\resizebox{\columnwidth}{!}{
    \begin{tabular}{lccc}
    \hline 
    Filament & Length $h^{-1}\mathrm{ Mpc}$ & DM particles & Density $\# /h^{-3}\mathrm{Mpc^3}$ \\
    \hline
    1  & 15.33 & 801  & 1.85 \\
    2  & 9.25  & 1844 & 7.05 \\
    3  & 8.15  & 1325 & 5.75 \\
    4  & 9.20  & 723  & 2.78 \\
    5  & 10.98 & 864  & 2.78 \\
    6  & 9.53  & 961  & 3.57 \\
    7  & 8.29  & 575  & 2.45 \\
    8  & 7.63  & 535  & 2.48 \\
    9  & 12.16 & 869  & 2.53 \\
    10 & 11.05 & 751  & 2.40 \\
    \hline
    \end{tabular}
}
\end{table}

\begin{figure}[h]
    \centering
    \includegraphics[width=0.85\linewidth]{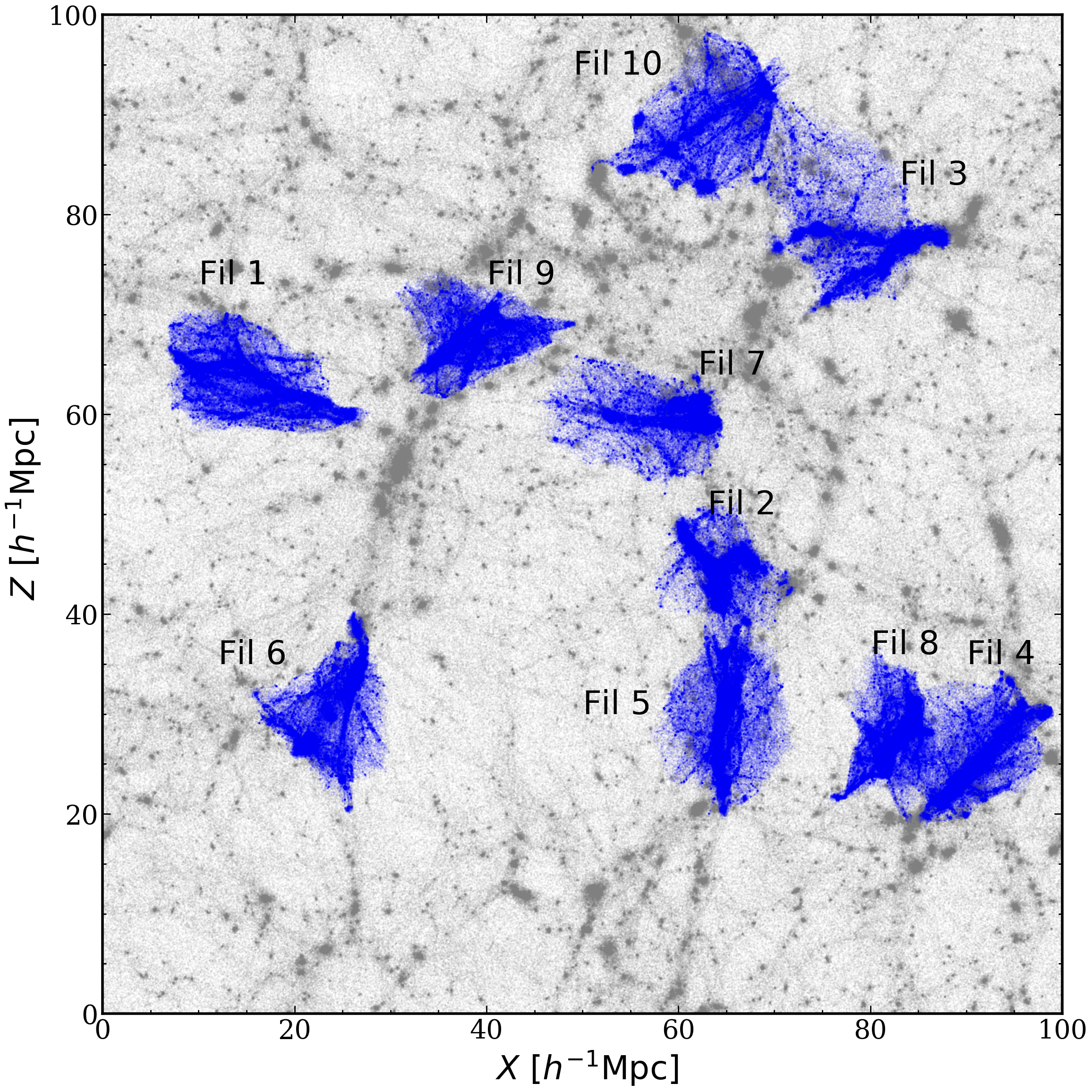}
    \caption{Side view of the zoomed-in region in which the ten filaments (in blue) at redshift 0  are located. Gray represents the particles of the level-7 initial simulation.}
    \label{filament_pro}
\end{figure}

To implement the zoom-in procedure, we used the particles associated with the filaments previously identified in the uniform-resolution (unigrid) level-7 simulation. For every filament, we traced the corresponding particles back to the initial conditions and defined an ellipsoidal Lagrangian region that is sufficiently large to enclose all particles that will eventually end up in the filament at the final snapshot. This ellipsoidal region is adopted as the high-resolution target volume for that filament, and a new set of multi-scale initial conditions is generated, in which the resolution is increased inside the zoom region, reaching a maximum refinement level of 11 (corresponding to a minimum DM particle mass of $m_{\mathrm{DM}} \simeq 1.01 \times 10^{7}~h^{-1}\mathrm{M_{\odot}}$) while the surrounding environment remains at a lower resolution. Each of these initial-condition sets is then used to run an independent zoom-in simulation.

The zoomed-in regions of the 10 filaments are shown in Fig.~\ref{filament_pro} (highlighted in blue), illustrating that the structures identified in the level-7 simulation remain unchanged at large scales, while the mass resolution within the refined regions is significantly enhanced. To require the curvature to be negligible, we used the zoom-in simulations. To quantify filament straightness, we used the filament spine extracted by the MBMS algorithm (blue points in Fig.~\ref{filament_cen}). We measured the perpendicular distance of each spine point to the adopted global axis and computed the radius $R_{90}$ containing $90\%$ of the spine points around it. The resulting values of $R_{90}/L$, where $L$ is the filament length, have a mean of 0.076 and a standard deviation of 0.047 across our sample. For reference, the filament shown in Fig.~\ref{filament_cen} has $R_{90}/L=0.040$, illustrating the high degree of straightness of the selected filaments. These values confirm that the adopted filaments are approximately straight despite residual variations in thickness and morphology.

The primary analysis was performed at $z=0$, while the evolutionary study considers snapshots in the range $1 \gtrsim z \gtrsim 0$. This interval is chosen because filaments are already well defined by $z=1$, allowing us to trace their influence, and because it probes a regime where nonlinear processes are expected to be important. Specifically, our analysis utilizes 12 full-particle snapshots equally spaced in scale factor between $z=1$ and $z=0$. Throughout this work, all the analysis is restricted to the highest-resolution (level-11) regions of the zoom-in simulations. This approach allows us to robustly resolve low-mass halos within filaments while maintaining a self-consistent cosmological environment.
Our filament-tracking methodology requires continuous access to particle information to reconstruct and follow the filament spine from $z=1$ to $z=0$. Restricting the analysis to high-resolution zoom-in regions provides a more manageable sample. Furthermore, restricting the analysis to 10 comparable, straight filaments provides a controlled environment in which mass-dependent alignment evolution can be studied while minimizing the additional complexity introduced by nodes and highly connected regions.

\subsection{Filament across redshifts} 
\label{filament_id}

To reconstruct our filament distances, we identified filaments at the highest redshift considered in our analysis, which allowed us to trace the densest and dynamically dominant regions of the structure and ensured a coherent definition of the filament backbone.
This analysis utilizes again the previously mentioned set of tools 1DREAM. Our reconstruction strategy follows a retrograde–anterograde tracking scheme. We defined the filament geometry at redshift $z = 1$, where the structure is already well established, by applying the full four-step pipeline (LAAT, MBMS, DimIndex, and Crawling) described in Sect.~\ref{zoomin}. This procedure yields a continuous and topologically robust definition of the filament spine at $z=1$. We then tracked the particles belonging to this spine across all snapshots down to $z=0$, re-applying MBMS at each step to follow the evolution of the filament’s physical backbone.

\begin{figure}[h]
    \centering
    \includegraphics[width=0.85\linewidth]{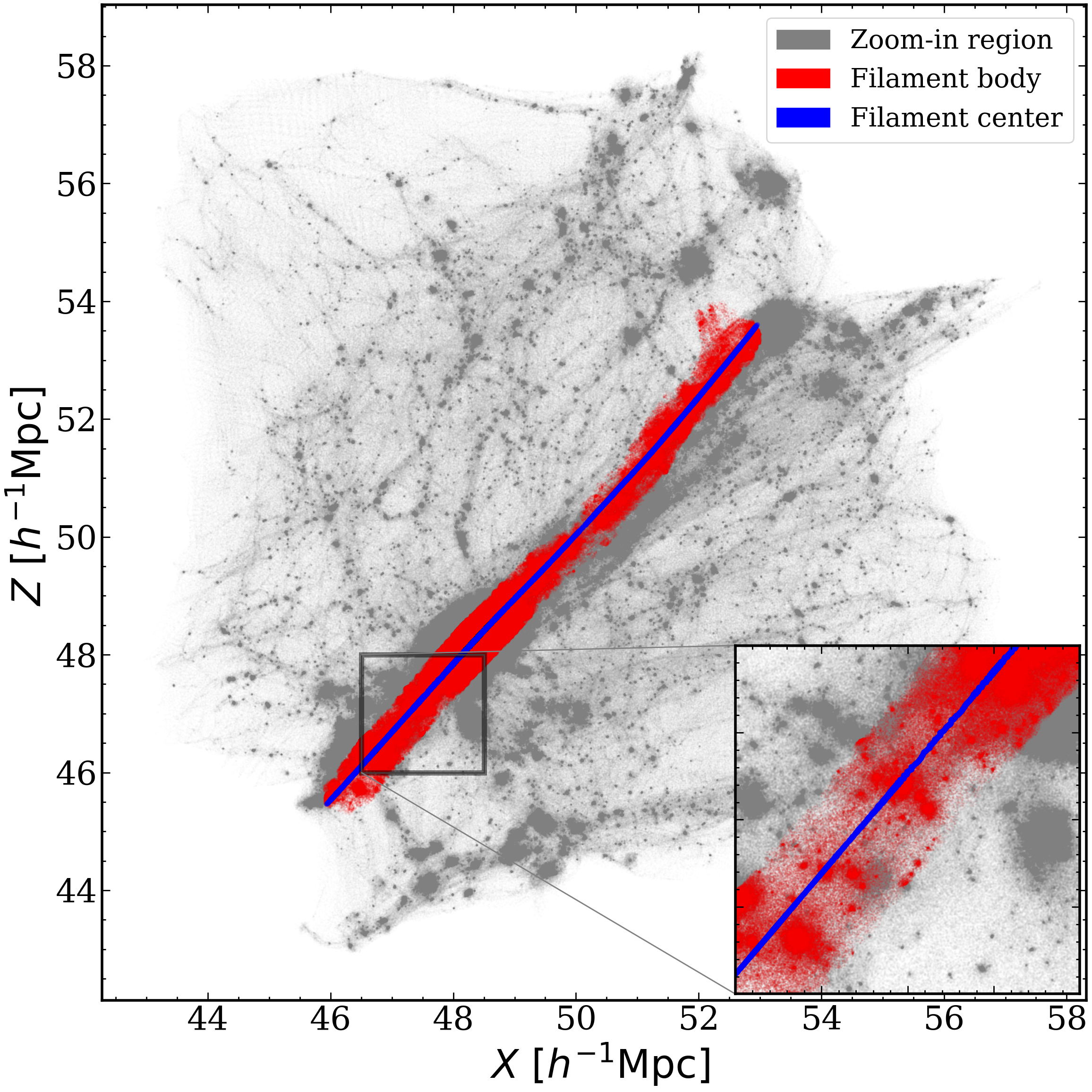}
    \caption{Side view of the zoomed-in region corresponding to filament 4 at redshift 1. Gray represents the particles of the level-11 zoom-in region, red shows the filament body detected by Crawling, and blue indicates the particles collapsed to the filament center by MBMS.}
    \label{filament_cen}
\end{figure}

For each halo identified in Sect.~\ref{halodata}, we defined its distance ($d$) to the filament as the minimum Euclidean distance between the halo’s center of mass and the reconstructed filament spine at the corresponding redshift. This provides a consistent, time-resolved metric of the halo's proximity to the deep potential well of the filament. Figure~\ref{filament_cen} illustrates the different regions identified for filament 4 at $z=1$.

\subsection{Halo properties}
\label{halodata}

For halo identification, we used the AMIGA Halo Finder (AHF) code \citep{AHF}, which identifies gravitationally bound structures in N-body simulations using the friends-of-friends method \citep{1982ApJ...257..423H}.
Due to the high resolution achieved by the zoom-in simulations, AHF is able to resolve a large number of low-mass DM halos. For this work, we adopted a  lower mass limit of $M_\mathrm{h} \geq10^{9.5}~h^{-1}\mathrm{M_\odot}$, which corresponds to a minimum of 311 DM particles per halo. This choice helped avoid resolution artifacts. 

Halo DM particles may not be strongly bound, but the angular momentum from each individual particle contributes to the total halo spin, defined as the sum over the angular momentum of all the DM particles that compose the halo:
\begin{equation}
\label{angular_momentum}
\textbf{J}= \sum_{n=1}^{N} m_n(\textbf{r}_n \times \textbf{v}_n),
\end{equation}
where $\textbf{r}_n$ and $\textbf{v}_n$ denote the position and velocity of the $n$-th particle relative to the halo center of mass, and $m_n$ is its mass. Commonly, the magnitude of the angular momentum, $J = |\textbf{J}|$, is used to quantify halo rotation through the spin parameter $\lambda$, first introduced by \citet{Peebles1969}. This parameter measures the relative importance of rotational support: values close to unity correspond to rotation-supported halos, while values near zero indicate pressure- or dispersion-supported systems \citep{1993sfu..book.....P}. In this work, we adopted the spin parameter definition of \citet{Bullock_2001}, denoted $\lambda'$. Defined within a virialized sphere of radius $R_{\rm vir}$ enclosing a mass $M_{\rm vir}$, with $V_{\rm vir}$ the circular velocity at $R_{\rm vir}$, the spin parameter is given by
\begin{equation}
\lambda'= \frac{J}{\sqrt{2}M_{\rm vir}V_{\rm vir} R_{\rm vir}}.
\end{equation}

 Throughout this work, we adopted the AHF virial quantities listed above.  Dark matter halos are often approximated as spherical systems for simplicity. However, N-body simulations and observational studies show that they typically deviate from spherical symmetry and are better described as triaxial ellipsoids \citep[e.g.,][]{2002ApJ...574..538J,Bailin_2005,2005ApJ...629..781K,2006MNRAS.367.1781A,2007MNRAS.377...50H,2011MNRAS.416.1377V,2015MNRAS.449.3171B,2016MNRAS.462..663B,2019MNRAS.490.4877P}. To characterize halo shape, we used the reduced moment of inertia tensor, $\tilde{I}_{ij}$, a $3 \times 3$ symmetric matrix defined as
\begin{equation}
    \tilde{I}_{ij} = \sum_{k=1}^N \frac{m_k r_{k,i}r_{k,j}}{r_k^2},
\end{equation}
where $r_k$ is the distance of the $k$-th particle from the halo center. This tensor can be diagonalized to calculate its eigenvalues, $s_\mathrm{a} \geq s_\mathrm{b} \geq s_\mathrm{c}$, and eigenvectors $\textbf{E}_\mathrm{a}, \textbf{E}_\mathrm{b}$, and $\textbf{E}_\mathrm{c}$, which denote the major, intermediate, and minor axes of the halo, respectively. We described the shape of the halos using the triaxiality parameter, $T$, defined as

\begin{equation}
    \label{triaxial_p}
    T=\frac{1-q^2}{1-s^2},
\end{equation}
in terms of the axis ratios $q = \mathrm{b}/\mathrm{a}$ and $s = \mathrm{c}/\mathrm{a}$, where $\mathrm{a} = \sqrt{s_\mathrm{a}}$, $\mathrm{b} = \sqrt{s_\mathrm{b}}$, and $\mathrm{c} = \sqrt{s_\mathrm{c}}$. This parameter allows halos to be classified into three regimes: when $T \rightarrow 0$ they tend to be oblate, for intermediate values they are triaxial, and when $T \rightarrow 1$ they are prolate.

To study the evolution of halos within the filament, we used AHF together with two complementary utilities: \texttt{MergerTree}, which traces the merger history of halos across snapshots, and \texttt{ahfHaloHistory}, which links the \texttt{MergerTree} output and generates individual files describing the evolution of each halo, reporting the properties of its most massive progenitor.

\section{Results}
\label{results}
\subsection{Spin and shape alignments at z=0}

In this section we focus on the alignment of halos with respect to their host filaments at $z=0$, where we considered only central halos that are enclosed within a cylindrical region of radius $3~h^{-1}\mathrm{Mpc}$ around the filament axis. This selection results in a total of 5,297 DM halos.

 \begin{figure}[h!]
    \centering
    \includegraphics[width=0.9\linewidth]{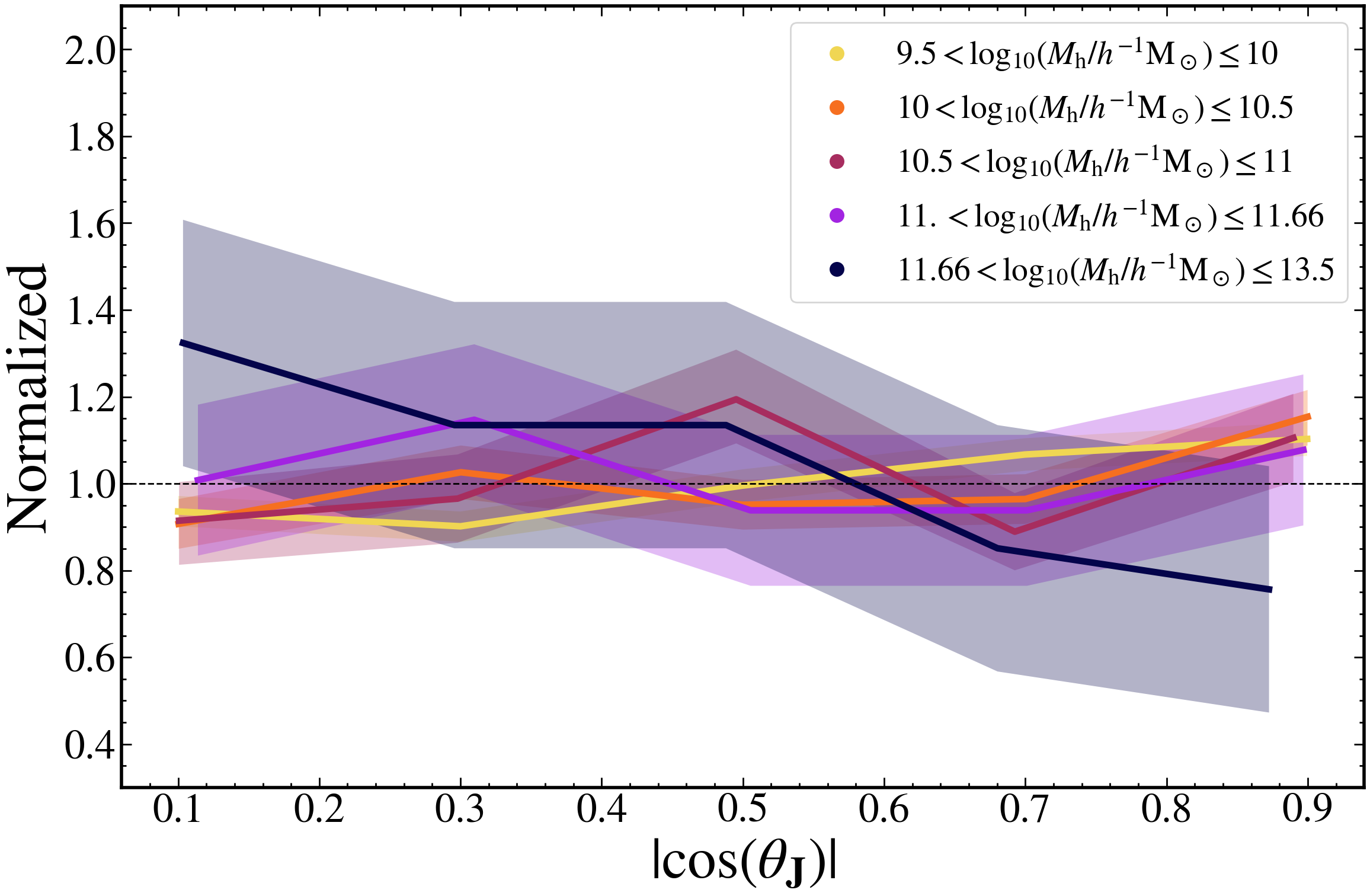}
    \caption{Stack of the spin--filament alignment distributions for the sample of ten filaments. The histograms are normalized to unity and separated into different mass bins, as indicated by the legend. The shaded regions represent the 1$\sigma$ uncertainty estimated via bootstrap resampling.}
    \label{global_spin_mass_S}
\end{figure}

\begin{figure*}[h]
  \centering
  \includegraphics[width=0.9\textwidth]{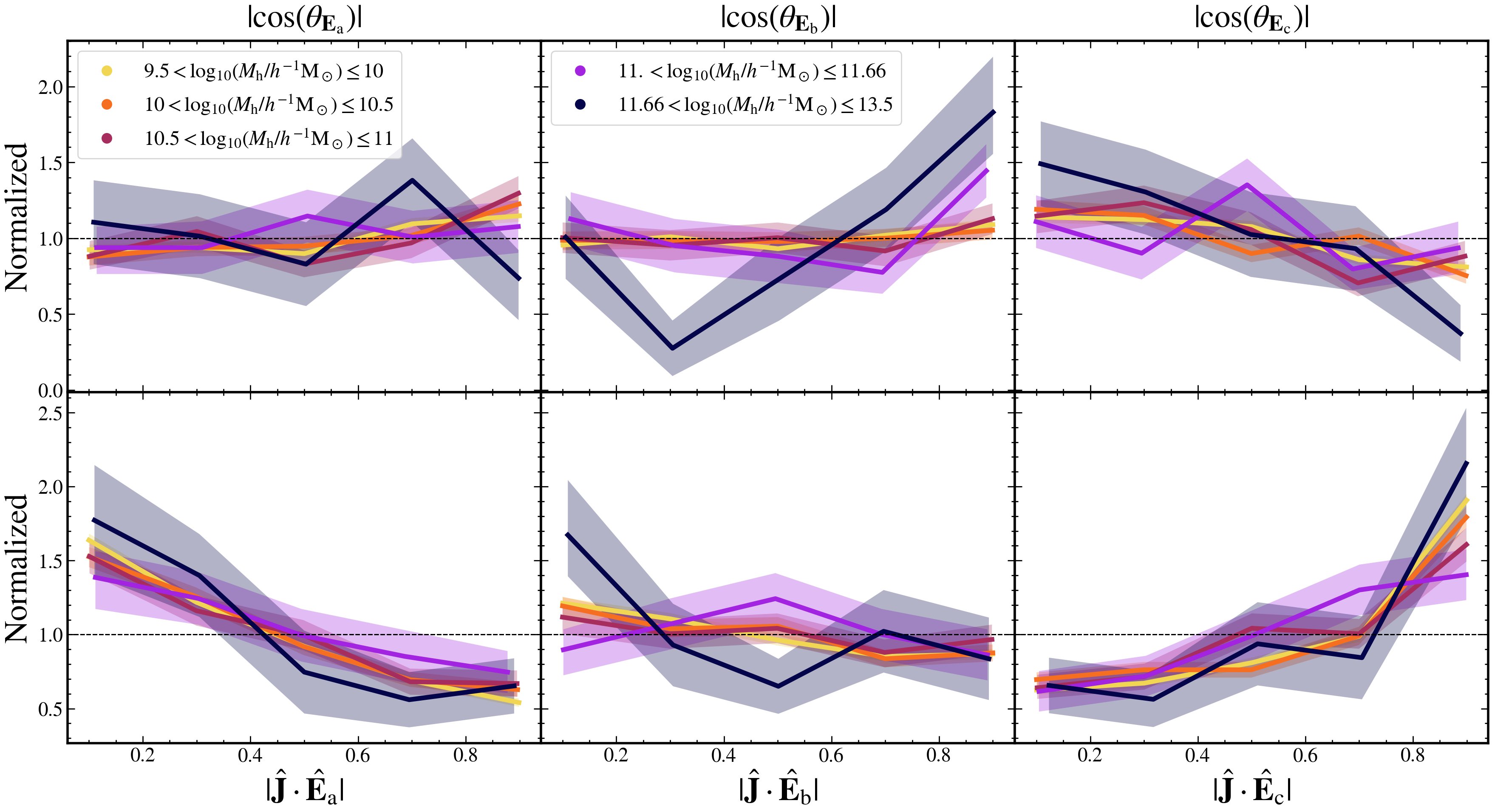}
  \caption{Similar to Fig.~\ref{global_spin_mass_S} but for the alignments related to the components of the inertia tensor. \textit{Top three panels}: Alignment between the filament direction and: the major axis, the second major axis, and the minor axis (denoted \(\textbf{E}_\mathrm{a}, \textbf{E}_\mathrm{b}\), and \(\textbf{E}_\mathrm{c},\) respectively). In each panel, halos are separated into different mass bins. \textit{Bottom three panels}: Dot product between the spin vector, \(\hat{\textbf{J}}\), and each of the inertia tensor axes, using the same mass binning as in the top panels. The shaded regions represent the 1$\sigma$ uncertainty estimated via bootstrap resampling.}
  \label{alignments_RI}
\end{figure*}

To quantify alignments, we defined $\theta$ as the angle between two vectors and computed $\cos\theta$ from the absolute value of their normalized dot product, restricting $\theta$ to the range $[0,\pi/2]$ since no preferred filament direction is assumed \citep[e.g.,][]{2014MNRAS.440L..46A, 10.1093/mnras/sty2270}. We then evaluated the alignment between (i) the halo spin and the filament axis (defined as the vector connecting the ends of the filament spine), $\cos(\theta_\mathbf{J})=|\hat{\mathbf{t}}_{\rm fil}\cdot\hat{\mathbf{J}}|$; (ii) the halo principal axes and the filament, $\cos(\theta_{\mathbf{E}_\text{i}})=|\hat{\mathbf{t}}_{\rm fil}\cdot\hat{\mathbf{E}}_\text{i}|$, where i=(a,b,c) denotes the major, intermediate, and minor principal axes, respectively; and (iii) the halo spin and shape, $|\hat{\mathbf{J}}\cdot\hat{\mathbf{E}}_\text{i}|$.

We stacked all the alignment values of halos from the ten filaments that make up our sample. To properly interpret the behavior, it is crucial to separate the sample by halo mass. The distribution of $|\cos\theta_j|$ for all halos located within our sample filaments at redshift 0 is shown in Fig.~\ref{global_spin_mass_S} in different halo mass bins, where the number of halos in each mass bin is normalized so that the area under the curve equals 1. 
To assess the statistical significance of the alignment signals, we estimated the uncertainties using a bootstrap resampling technique with $5\,000$ realizations. We note that this halo-level approach assumes independence and may not fully capture filament-to-filament differences and cosmic variance. In addition, we performed a Kolmogorov--Smirnov (KS) test, comparing the measured distributions with a uniform distribution. For the lowest-mass bin $(9.5 < \log_{10}(M_\mathrm{h}/h^{-1}\mathrm{M_\odot}) \le 10.0)$ we obtain $D = 0.0414$ and $p = 8.46 \times 10^{-5}$, which indicates a statistically significant deviation from isotropy. In contrast, the highest-mass bin $(11.7 < \log_{10}(M_\mathrm{h}/h^{-1}\mathrm{M_\odot}) \le 13.5)$ yields $D = 0.1335$ and $p = 2.57 \times 10^{-1}$, which may be due to the smaller number of halos in this mass range.

Consistent with previous results reported in the literature \citep[e.g.,][]{10.1093/mnras/stv1570,10.1093/mnras/sty2270,2018MNRAS.481.4753C,2020MNRAS.493..362K,10.1093/mnras/stab411}, we find a clear mass-dependent trend in the alignment signal. Low-mass halos tend to align parallel to the filament axis, whereas the most massive halos exhibit a preferential perpendicular orientation. Although the signal in the high-mass regime is not statistically significant according to the KS test, it follows the same overall trend, albeit with larger uncertainties due to limited statistics.

We identify a transition between these two regimes occurring near $\log_{10}(M_\mathrm{h}/h^{-1}\mathrm{M_\odot}) \sim 12$, where perpendicular alignments begin to dominate. This transition mass is consistent with previous findings \citep[e.g.,][]{Codis_2012,10.1093/mnras/sty2270,2019MNRAS.485.5244L}.

So far, our analysis has focused on the orientation of the spin vector relative to the host filament. We also investigated the alignment of the halo's shape. Previous studies have established that the minor axes of DM halos tend to be oriented perpendicular to the filaments or walls of the cosmic web \citep{aragon-calvo,10.1111/j.1365-2966.2006.11318.x,10.1093/mnras/stw1247,10.1093/mnras/sty2270,10.1093/mnras/stab451}, indicating a strong correlation between the halo's geometry and its large-scale environment.

In the top three panels of Fig.~\ref{alignments_RI}, we present the alignment between these three principal axes and the filament vector. Two main results emerge from this analysis. First, unlike the spin vector, the orientation of the inertia tensor components does not exhibit a directional inversion (or ``shape-flip'') as a function of halo mass. However, as we explore in the next section, the strength of this alignment does depend on mass. Second, the major and intermediate axes tend to align parallel to the filament, whereas the minor axis is preferentially oriented perpendicular to it across all mass bins \citep{aragon-calvo,10.1111/j.1365-2966.2006.11318.x,10.1093/mnras/stw1247,10.1093/mnras/sty2270,10.1093/mnras/stab451}.

Additionally, we examined the internal alignment between the spin vector and the halo's shape. The bottom three panels of Fig.~\ref{alignments_RI} display the angle between the spin vector and each principal axis. We observe a trend inverse to that of the filament alignment: the spin vector tends to be perpendicular to the major and intermediate axes, while showing a strong parallel alignment with the minor axis. Physically, this indicates that these halos, which tend to be prolate, rotate preferentially around their shortest axis.
To quantify the statistical significance of these trends, we again performed a KS test for each mass bin. As in the spin--filament alignment discussed previously, the lowest-mass bin shows the strongest statistical significance, with $p$-values well below 0.05 for most measurements. However, unlike the spin case, several of the alignments in the highest mass bin also yield statistically significant results. This suggests that shape-related alignments remain comparatively strong even at high masses.

\subsection{Alignment evolution}
\label{aligment_evo}
Halos acquire their initial angular momentum through tidal torques during the early linear regime, as described by the TTT. However, several studies (e.g., \citealt{2025MNRAS.543.2222L}) have shown that deviations from TTT predictions already emerge at very high redshift ($z \sim 10$), even before the onset of strong nonlinear evolution. Motivated by this, we focused on the nonlinear regime, where TTT no longer holds, analyzing the spin evolution of our sample of halos in the range $0 \leq z \leq 1$. To investigate how the filament affects the evolution of halo spin, we restricted our sample to halos that can be traced back to $z = 1$, a redshift at which filaments can be clearly identified and nonlinear processes are already at play. This allowed us to follow their evolution down to $z=0$. This selection resulted in 4,797 halos.

\begin{figure}[h!]
    \centering
    \includegraphics[width=0.9\linewidth]{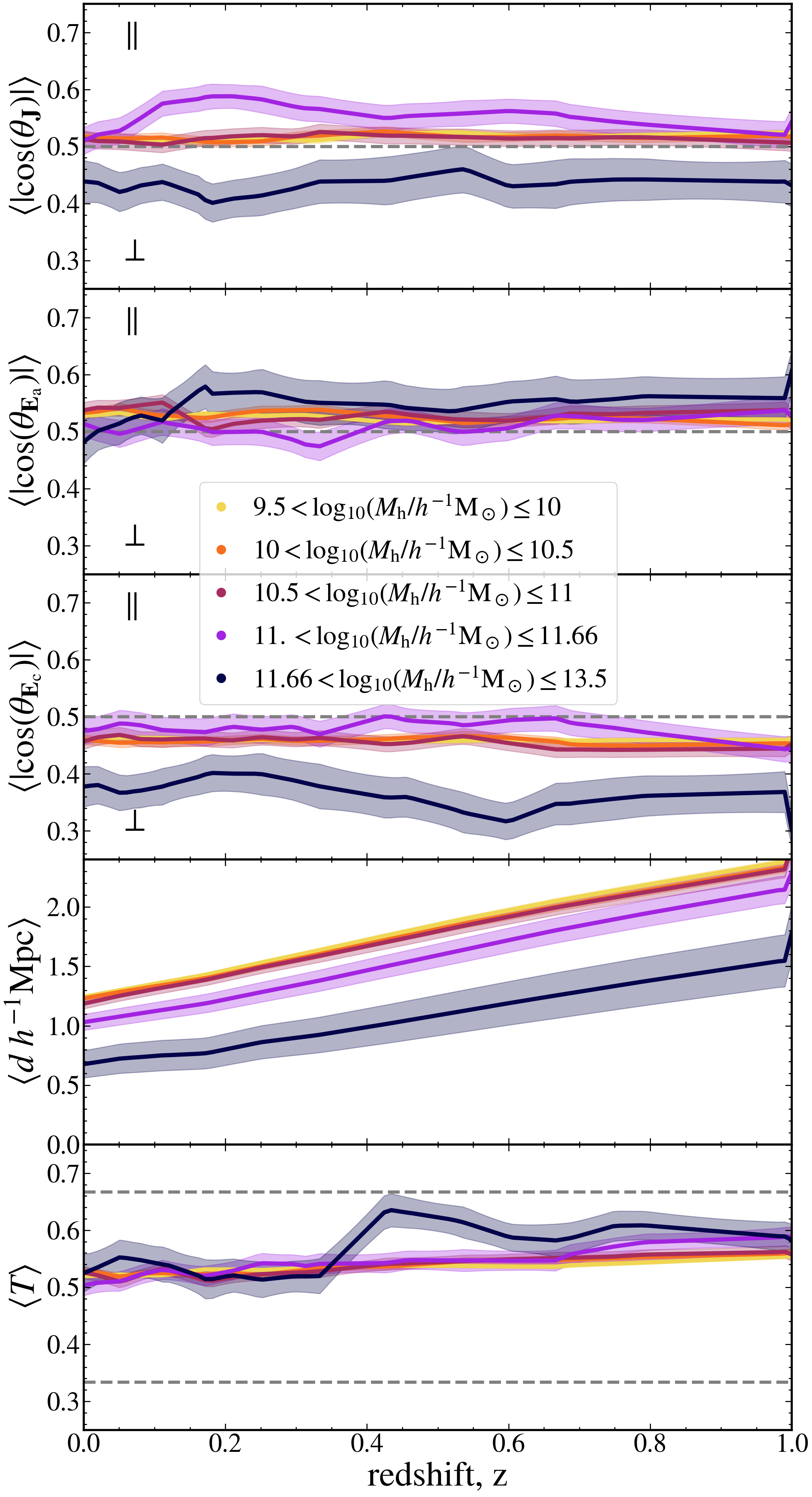}
    \caption{Average evolution of the halos belonging to the filaments at $z=0$ ($r \leq 3 ~h^{-1}\mathrm{Mpc}$), colored according to the final mass at $z=0$, with shaded regions indicating the standard error of the mean. \textit{Top panel}: Spin alignment with respect to the filament. \textit{Next two panels}: Shape alignment of the major and minor axes with respect to the filament. \textit{Fourth panel}: Distance from the filament. \textit{Bottom panel}: Triaxiality parameter ($T$).}
    \label{curves_avg_cumm_cen}
\end{figure}

Figure~\ref{curves_avg_cumm_cen} shows the interpolated evolutionary curves of all halos contained in all filaments, grouped by their mass at redshift $z=0$.
Halos in the highest $z=0$ mass bin tend to exhibit a stronger perpendicular alignment with the filament. Toward lower masses, this behavior gradually shifts toward a more parallel configuration, although the trend becomes less pronounced in the lowest-mass bin, in agreement with what is observed in Fig.~\ref{global_spin_mass_S}.
Another key result shown in this figure is the remarkable stability of this signal since $z \simeq 1$. Despite halos progressively approaching the filament spine, their orientation remains nearly unchanged over time, indicating that the alignment observed at $z=0$ was largely established at earlier epochs.

We also tracked the evolution of the orientation of the inertia tensor's components relative to the filament, focusing primarily on the major and minor axes. Similar to the spin alignment, these tracks do not exhibit significant variations with redshift, except for the major axis in the highest mass bin, which remains relatively constant over most of its evolution, but exhibits a change in its alignment at low redshift. While the minor axis maintains the perpendicular orientation established at high redshift, showing an almost unperturbed evolution, the intermediate axis (not shown) remains statistically neutral (with alignment values around 0.5) throughout cosmic time.

Regarding the spatial distribution (fourth panel), we find a clear mass dependence. At $z=0$, the most massive halos are preferentially located closer to the filament spine than low-mass halos. Since the mass bins are defined at $z=0$, their progenitors are less massive at higher redshifts, which should be kept in mind when interpreting the evolutionary trends. We also observe a mass-dependent trend in halo shape described by the triaxiality parameter (bottom panel). Halos in all mass bins (imposed at $z=0$) evolve to become more spherical as cosmic time progresses. At $z=1$ we find a clearer trend where massive halos exhibit more prolate shapes while low-mass halos are more spherical. This mass-dependent shape evolution is consistent with previous findings \citep[e.g.,][]{2011MNRAS.416..248R}. Finally, as redshift approaches $z=1$, most curves tend to converge toward a common region, indicating a possible settling of these properties during halo evolution.

We also compared the mean alignment histories of halos associated with filaments of different densities (not shown). We find no significant differences, with all filaments exhibiting similar spin filament alignment distributions.

\subsection{Drivers of spin alignment}
\begin{figure*}[h!]
  \centering
  \includegraphics[width=0.95\textwidth]{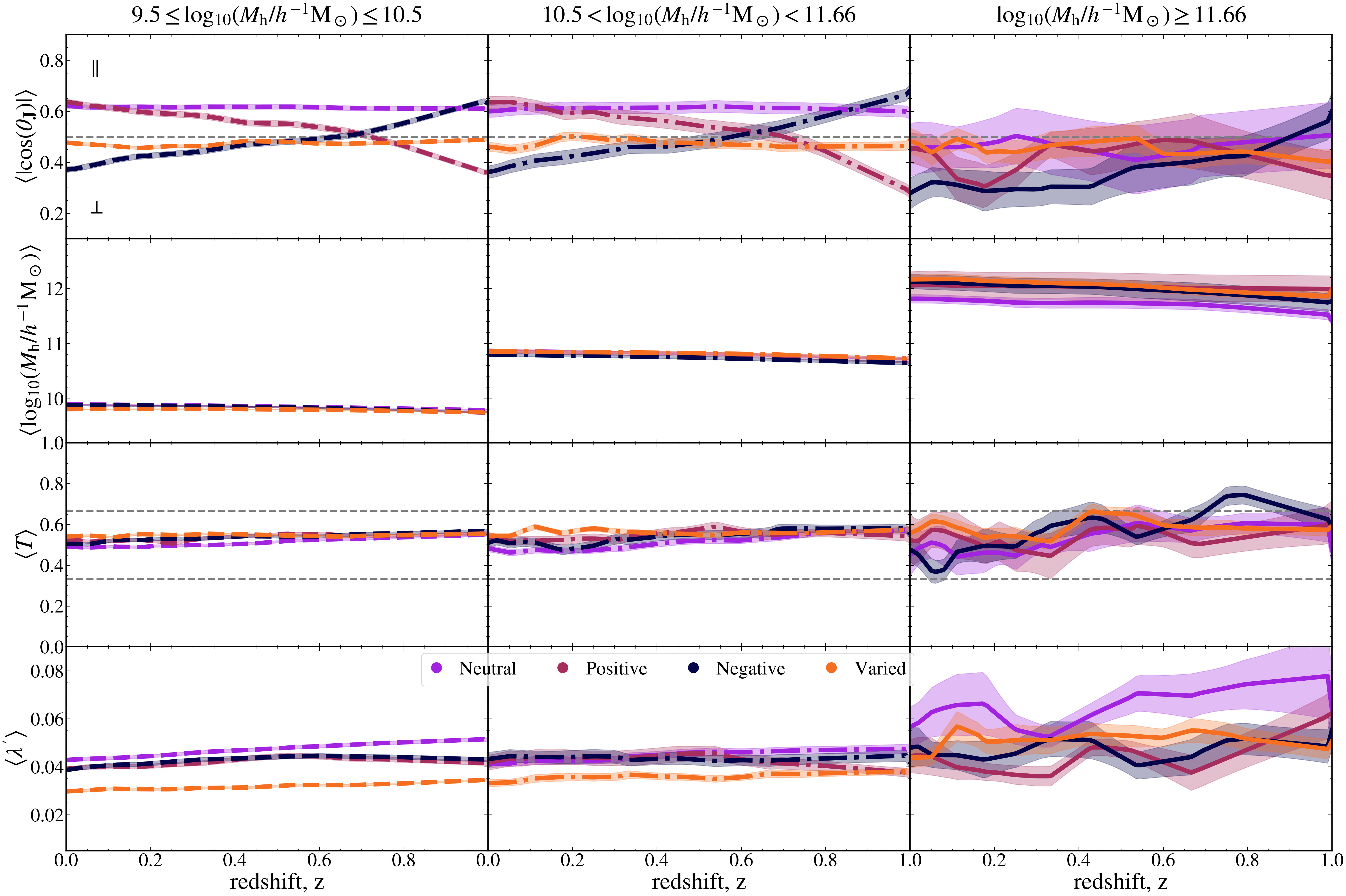}
  \caption{Similar to Fig.~\ref{curves_avg_cumm_cen} but for the average evolution of halos grouped according to the type of alignment evolution curve they exhibit: black (neutral) for halos that do not experience significant changes, yellow (positive) for halos that become more parallel over time, red (negative) for halos showing a decrease in alignment, and orange (varied) for halos whose spin evolution does not fit into the previous three categories. Each column represents a different mass bin, arranged from left to right as follows: low mass ($9.5 \leq \log_{10}(M_\mathrm{h}/h^{-1}\mathrm{M_\odot}) \leq 10$), intermediate mass ($10.5 < \log_{10}(M_\mathrm{h}/h^{-1}\mathrm{M_\odot}) < 11.66$), and high mass ($\log_{10}(M_\mathrm{h}/h^{-1}\mathrm{M_\odot}) \geq 11.66$). \textit{Top to bottom}: Spin--filament alignment, halo mass, triaxiality parameter, and spin parameter.}
  \label{alignment_curve}
\end{figure*}

\begin{figure*}[h]
  \centering
  \includegraphics[width=0.95\textwidth]{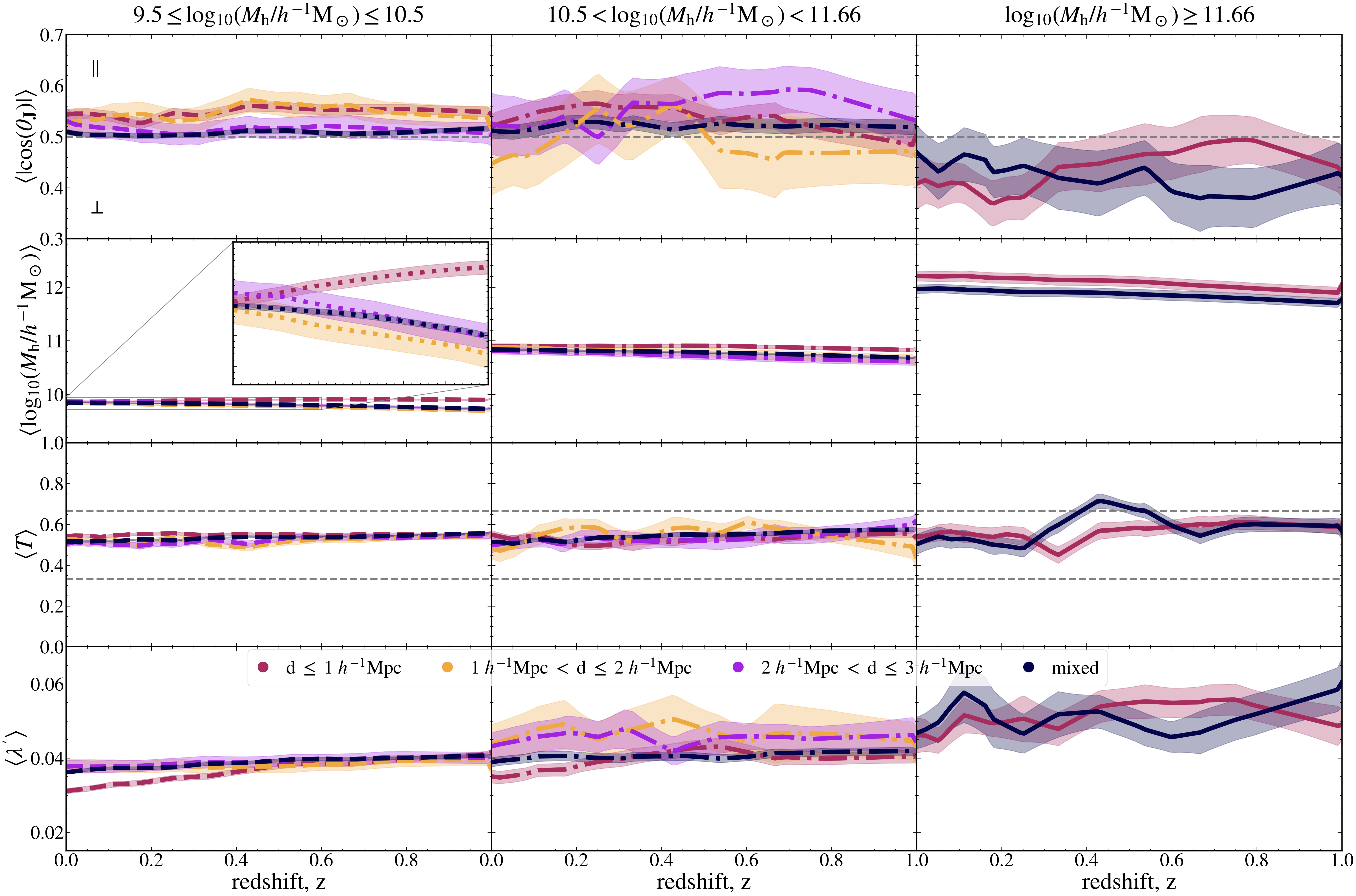}
\caption{Same as Fig.~\ref{alignment_curve} but grouping halos according to their environmental history instead of their alignment evolution. The curves are color-coded according to the distance $d$ (in $\mathrm{h^{-1}Mpc}$) from the halos to the filament center, as indicated in the legend. The mixed category includes halos that traverse multiple environments throughout their evolution.}
  \label{environment}
\end{figure*}

\begin{figure*}[h]
  \centering
  \includegraphics[width=0.95\textwidth]{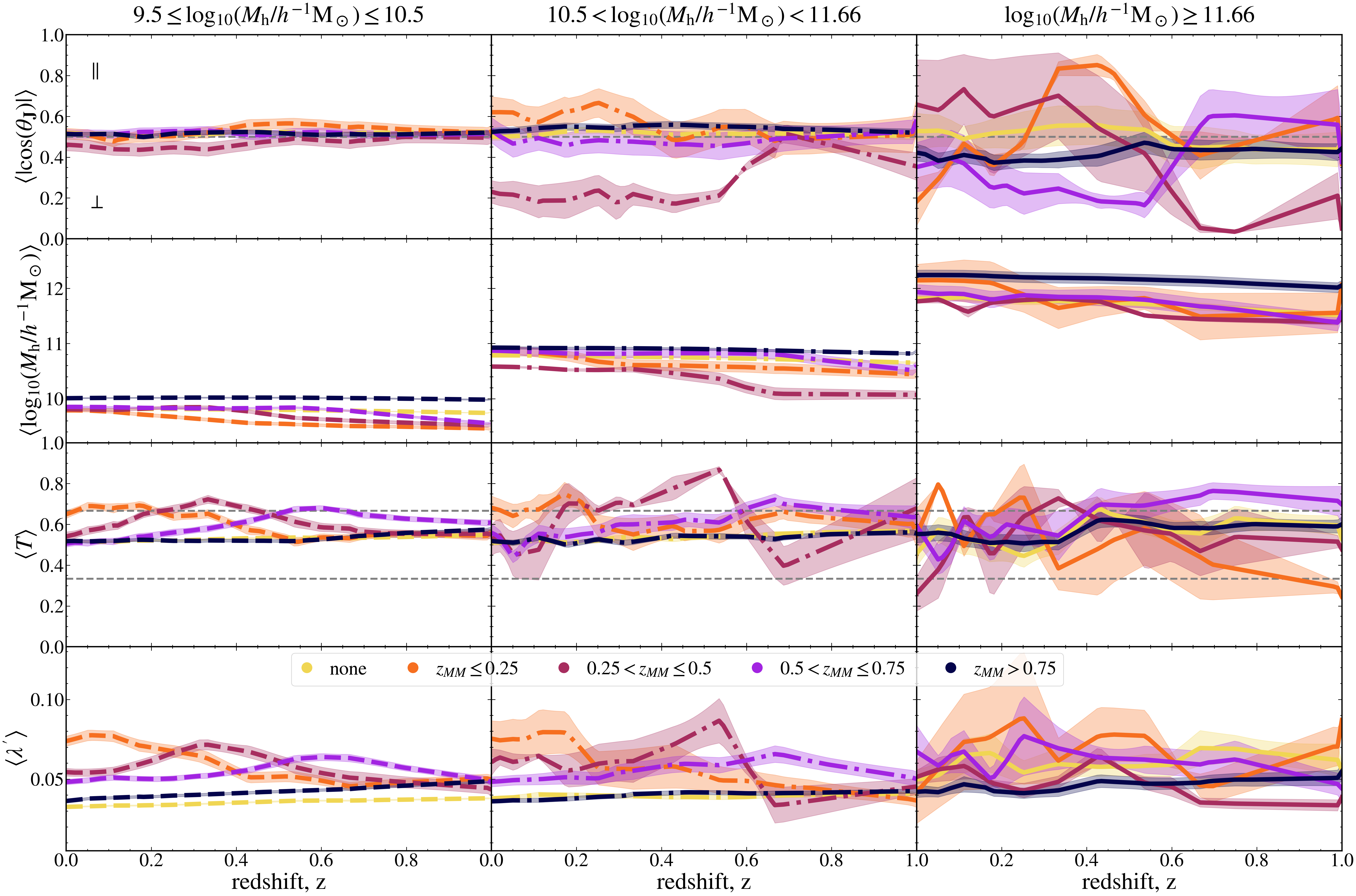}
\caption{Same as Fig.~\ref{alignment_curve} but for the subsets based on the time of the LMM. The curves are classified according to the redshift of the last event ($z_{\mathrm{MM}}$), as indicated in the legend. The category labeled ``none'' corresponds to halos that have not undergone a MM.}
  \label{TSLMM}
\end{figure*}

We categorized the halo population based on three distinct drivers: the history of spin--filament alignment, the distance to the host filament, and the time of the last major merger (LMM). Figures~\ref{alignment_curve}, \ref{environment}, and \ref{TSLMM} illustrate the average evolution of these subsets. Given that halo mass plays a decisive role in the acquisition and redistribution of angular momentum, we further subdivided each subset into three mass bins defined at $z=0$ to disentangle its effects: low ($9.5 \leq \log_{10}(M_\mathrm{h}/h^{-1}\mathrm{M_\odot}) \leq 10.5$), intermediate ($10.5 < \log_{10}(M_\mathrm{h}/h^{-1}\mathrm{M_\odot}) < 11.66$), and high mass ($\log_{10}(M_\mathrm{h}/h^{-1}\mathrm{M_\odot}) \geq 11.66$).

\subsubsection{Subsampling by alignment evolution}
\label{subalignment}


To classify halo alignment evolution, we implemented an automated algorithm that analyzes snapshot-to-snapshot variations in the $|\cos(\theta_\mathrm{J})|$ tracks. To avoid classifying minor fluctuations as genuine reorientations, we adopted a threshold of $\Delta|\cos(\theta_\mathrm{J})|=0.2$ ($20\%$ of the parameter range). Based on this criterion (Fig.~\ref{alignment_curve}), tracks are classified as neutral if their peak-to-peak variation remains below the threshold; positive or negative if they exhibit a net increase or decrease larger than 0.2, respectively; and varied if they show both positive and negative variations exceeding the threshold.

\begin{table}[h!]
\small  
\centering
\caption{Number of halos by alignment evolution curve and mass.}
\begin{tabular}{lccc}
\hline
 &Low-mass & Intermediate-mass & High-mass \\
\hline
 Neutral   & 1013 & 169 & 7 \\
 Positive  & 526  & 84  & 6 \\
 Negative  & 574  & 89  & 10 \\
 Varied    & 2060 & 227 & 32\\
\hline
\label{laign_info}
\end{tabular}
\end{table}

The number of halos in each category is listed in Table~\ref{laign_info}. Figure~\ref{alignment_curve} shows the average evolution of several halo properties. Examining the alignment tracks, we clearly distinguish the characteristic behaviors of the neutral, positive, negative, and varied categories. We observe a mass dependence: the downward shift of the curves becomes evident in the highest-mass bin, indicating that these halos tend to evolve toward perpendicular orientations relative to the filament. Despite the limited statistics, the downward trend remains consistent within the shaded uncertainty region. The neutral curve generally lies slightly above the others in most mass bins, particularly in the two lower-mass bins. This tendency is more evident at certain redshifts and contributes, together with the positive curve, to the population of halos exhibiting parallel orientations relative to the filament. Conversely, the negative and varied curves drive the perpendicular alignment signal observed at $z=0$. The varied curves remain close to the random alignment value of 0.5, typically lying slightly below it. This indicates that halos with erratic alignment histories do not exhibit a strong preferred orientation on average.

The contrast between neutral and negative halos suggests two different evolutionary pathways. Neutral halos maintain nearly constant alignments and may preserve the orientation already established by $z=1$, whereas negative halos experience a progressive reorientation toward perpendicular configurations.

Regarding physical properties beyond alignment, such as the distance to the filament (not shown), we find no significant correlation with alignment type, except for the mass segregation effect discussed previously. Likewise, mass evolution follows the expected growth for each mass bin and shows no dependence on alignment type. A small difference appears in halo shape: halos with a neutral alignment history tend to be slightly more oblate ($T \to 0$), while those with varied histories show a mild tendency toward more prolate configurations ($T \to 1$).

For the spin parameter $\lambda'$, the trends are more distinguishable than for the shape. In the low-mass bins, halos with a neutral alignment history tend to have slightly larger spin parameters compared to the other categories, while the varied halos show marginally lower values of $\lambda'$. This difference is less evident in the high-mass bin. 
In the two lowest-mass bins, these populations also tend to lie somewhat closer to the filament. Since this region is mildly denser and dynamically more complex, this spatial segregation may contribute to the lower spin values observed for the varied halos, although this interpretation remains tentative.

\subsubsection{Environmental dependence}
\label{sunenviron}

To investigate how different regions in the filament affect halo evolution, we classified halos based on their distance to the filament's center. Specifically, we defined three environmental categories: halos that remain consistently within 1 $h^{-1}\mathrm{Mpc}$, those located between 1 and 2 $h^{-1}\mathrm{Mpc}$, and those between 2 and 3 $h^{-1}\mathrm{Mpc}$ from the filament.

In Fig.~\ref{environment} we present the evolution of halos that remain in one of these distance-defined environments throughout their history. We also include an additional category labeled ``mixed,'' which contains halos that change environment over time and do not remain in a single region. The number of halos in each category is presented in Table~\ref{env_info}, together with the mean minimum distance reached by the mixed halos.

\begin{table}[h!]
\small
\centering
\caption{Number of halos by environment and mass.}
\label{env_info}

\begin{tabular}{lccc}
\hline
Environment & Low-mass & Intermediate-mass & High-mass \\
\hline
$d \leq 1\,h^{-1}\mathrm{Mpc}$ & 785 & 125 & 31 \\
$1$--$2\,h^{-1}\mathrm{Mpc}$ & 146 & 18 & 0 \\
$2$--$3\,h^{-1}\mathrm{Mpc}$ & 181 & 25 & 1 \\
\makecell[l]{Mixed \\ $\langle d_{\rm min}\rangle \approx 1.3\,h^{-1}\mathrm{Mpc}$}
& 3061 & 401 & 23 \\
\hline
\end{tabular}

\end{table}

The alignment evolves as expected for each mass bin when considering the distance. As indicated in Table~\ref{env_info}, most massive halos remain closer to the filament throughout their evolution, which is associated with a strong perpendicular alignment trend. In contrast, halos in the lowest-mass bin show alignments that are predominantly parallel to the filament, while those in the intermediate mass bin exhibit a qualitatively similar behavior, except in the 1–2~$h^{-1}\mathrm{Mpc}$ range, where no well-defined evolutionary trend is observed, although they tend to be perpendicular to the filament at $z=0$.

No further variations regarding distance are discussed, as this parameter was constrained by our selection criteria. Similar to Fig.~\ref{alignment_curve}, only small differences with mass selection are visible. In the zoomed panel, the lowest-mass bin shows a slight decrease in halos closest to the filament. However, this effect is minor and may be related to residual mass segregation within the bins. This is presumably because they are less strongly bound than massive halos, and interactions with the densest part of the filament lead to mass stripping. Additionally, the curves for the closest halos in the other two mass bins remain above the others, consistent with the fact that more massive halos tend to reside closer to the filament center, as expected from previous studies and as also shown in Fig.~\ref{curves_avg_cumm_cen}.

Regarding shape, there is no distinctive trend driven by the environment. Finally, for the spin parameter, a decrease for the halos closest to the filaments is evident in the low- and intermediate-mass bins, whereas the highest-mass bin does not show a comparable trend. In the two lowest-mass bins, the other two environments show higher spin values.

\subsubsection{Subsampling by last major merger}

Major mergers are widely known to produce significant changes in both galaxies and their host DM halos. In this work, a MM is identified by analyzing halos at all available redshifts and tracing their progenitors in the previous snapshot using the merger trees. The most massive progenitor is defined as the main progenitor, and the merger is classified as major when the combined mass of the remaining progenitors exceeds one-fourth of the mass of the main progenitor. In Fig.~\ref{TSLMM} we present a classification based on the time of the LMM, dividing the sample into five subsamples: halos whose last MM occurred at $z_{MM} \leq 0.25$, $0.25 < z_{MM} \leq 0.5$, $0.5 < z_{MM} \leq 0.75$, and $z_{MM} > 0.75$, as well as halos that have not undergone any MM. The distribution of these subsamples is listed in Table~\ref{TSLMM_info}.

\begin{table}[h]
\small 
\centering
\caption{Number of halos by LMM and mass.}
\begin{tabular}{lccc}
\hline
 &Low-mass & Intermediate-mass & High-mass \\
\hline
 $z_{MM} \leq 0.25$       & 105 & 12 & 2 \\
 $0.25 < z_{MM} \leq 0.5$ & 83  & 3  & 2 \\
 $0.5 < z_{MM} \leq 0.75$ & 182  & 22  & 5 \\
 $z_{MM} > 0.75$          & 705 & 243 & 34\\
 None                     & 3098 & 289 & 12\\
 \hline
\label{TSLMM_info}
\end{tabular}
\end{table}

 In the high-mass bin, the effect of the LMM is reflected in a change in alignment, with halos becoming either more parallel or more perpendicular to the filament. Although the uncertainty bands are larger, they still support the observed alignment trend. Halos that become perpendicular show a clearer connection with the merger time, whereas those that become parallel tend to evolve earlier and therefore do not exhibit an equally clear correlation. The subsample that exhibits an unperturbed evolution across all three mass bins corresponds to halos with $z_{MM} > 0.75$.

We also see a clear impact on mass growth: MMs produce an increase in mass at the time of the event, consistently across the three mass bins. The shape evolution becomes especially interesting, as MMs induce a temporary transition toward a more prolate configuration. Since these mergers typically occur along the major axis, the associated stretching modifies the halo’s shape. Finally, the spin parameter is also affected by the merger, showing a noticeable increase at the time of the event. This behavior can be interpreted as a consequence of the accretion of orbital angular momentum during the MM, coupled with a redistribution of the halo’s kinetic and potential energy. Such events induce transient changes in the halo's internal structure, temporarily enhancing the relative contribution of ordered angular momentum to the system. This interpretation is consistent with our findings that, for early mergers, the post-merger increase in the spin parameter subsequently decays and stabilizes, indicating that these merger-induced effects are largely transient.

\begin{figure}[h]
  \centering
  \includegraphics[width=0.9\linewidth]{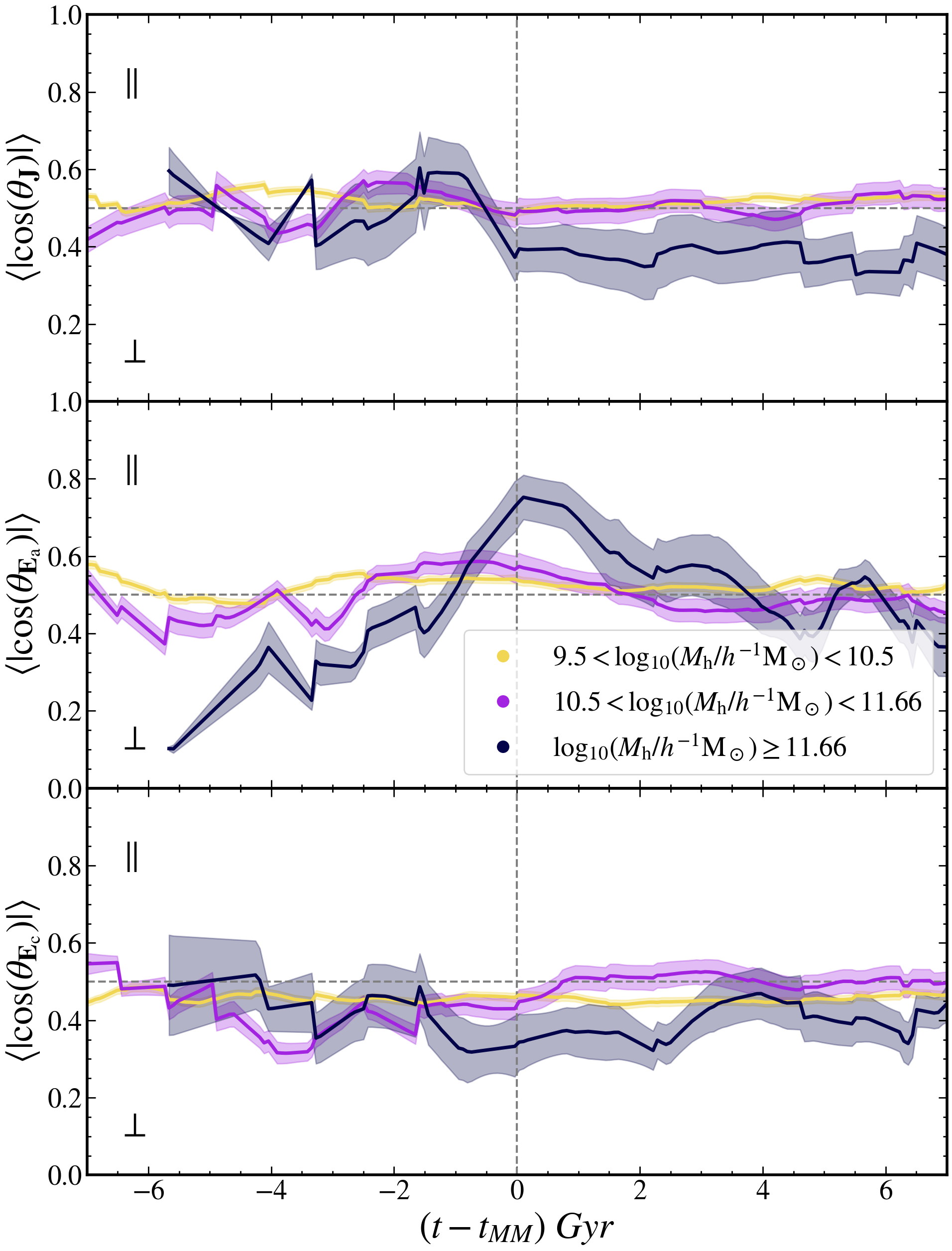}
  \caption{Average evolution of halos from the ten filaments, binned by mass, as a function of cosmic time relative to the LMM time. \textit{Top}: Evolution of the spin alignment. \textit{Middle}: Alignment of the major axis of the inertia tensor. \textit{Bottom}: Alignment of the minor axis of the inertia tensor.}
  \label{zmerger}
\end{figure}

Given that MMs have the strongest impact on alignment, we investigated this effect in greater detail. Retaining the same mass binning, we shifted from a static classification based on the time of the last merger to a dynamical analysis of the event itself. 
Figure~\ref{zmerger} shows the evolution of the alignment of both the spin vector and the principal axes of the inertia tensor with respect to the filament as a function of time relative to the merger epoch, for mergers occurring between $z=1$ and $z=0$. The time axis is defined such that the merger event occurs at $t=0$. This selection yields 360, 79, and 14 halos in the low-, intermediate-, and high-mass bins, respectively.

The results demonstrate that MMs induce a distinct reorientation, particularly strong for the most massive halos (albeit constrained by the small number of such events in our sample). For the spin vector, a decrease in alignment (i.e., a shift toward perpendicular orientation) becomes evident even prior to the merger event ($t < 0$), likely due to tidal pre-interaction. The strength of this signal diminishes with decreasing halo mass, becoming negligible for the lowest-mass bin. Regarding the inertia tensor, we observe a coherent response: the major axis alignment increases (becoming parallel to the filament) coincident with the spin becoming perpendicular. Conversely, the minor axis follows the spin's behavior, exhibiting a sharp drop in alignment. This trend is most pronounced in the high-mass bin.

\begin{figure}[h]
  \centering
  \includegraphics[width=0.9\linewidth]{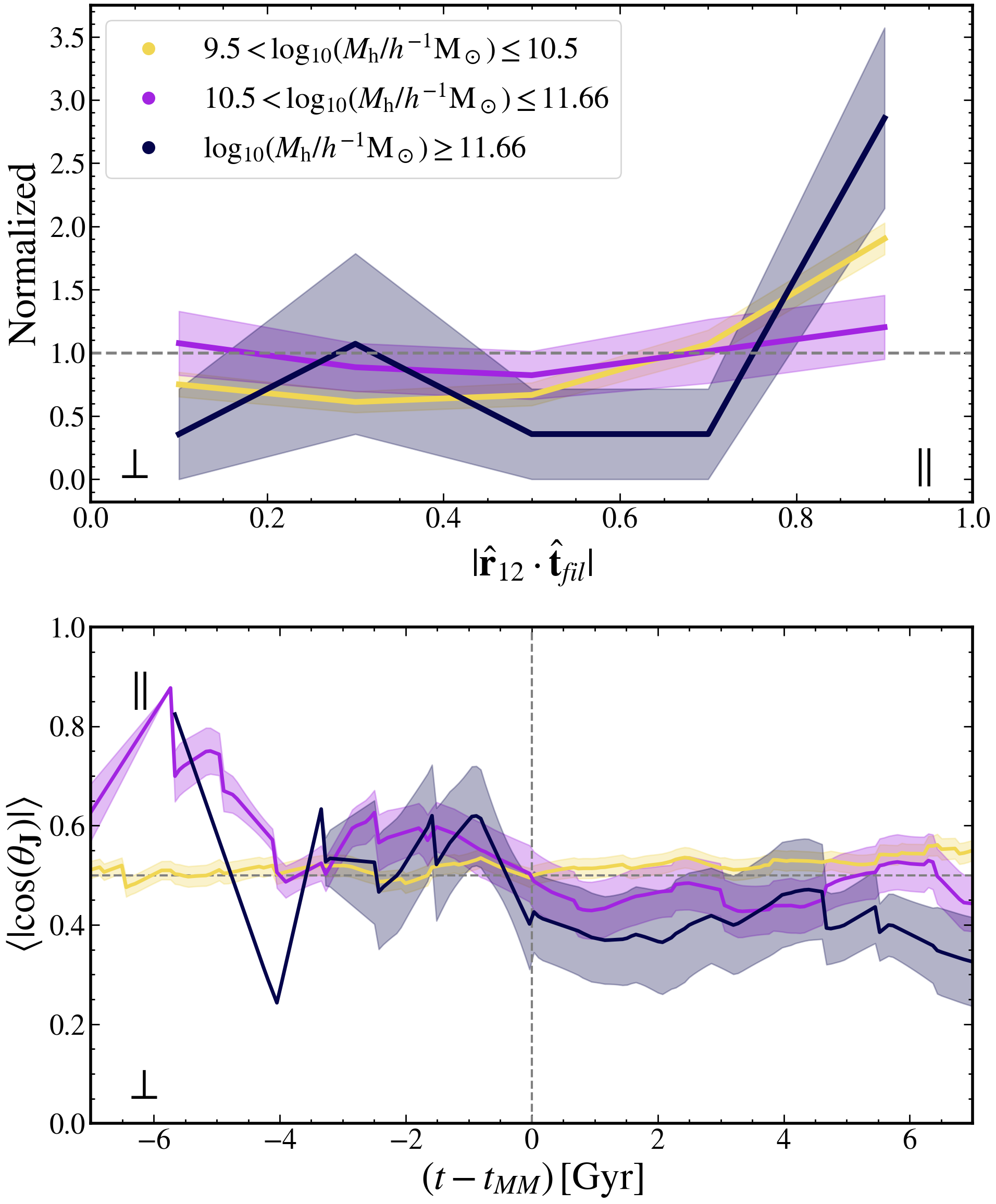}
\caption{\textit{Top}: Distribution of the alignment between the progenitor separation vector and the filament direction, evaluated at the time of the LMM for the same halo sample presented in Fig.~\ref{zmerger}. The probability density distributions are shown for different halo mass bins, as indicated in the legend. The shaded regions represent the $1\sigma$ uncertainties estimated through bootstrap resampling.
\textit{Bottom}: Average alignment evolution for the halo populations presented in Fig.~\ref{zmerger} but restricted to halos whose merger trajectory is strongly aligned with the filament, defined by a merger--filament alignment greater than 0.7 at the time of the LMM. The shaded region indicates the standard error of the mean. This selection isolates mergers occurring preferentially parallel to the filament direction.}
  \label{alignmer}
\end{figure}

As shown by \citet{10.1093/mnrasl/slu106}, mergers preferentially occur along the direction of cosmic filaments. Consequently, the orbital angular momentum of the merging pair is oriented perpendicular to the filament axis. Figure~\ref{alignmer} presents the normalized distribution of $|\hat{\mathbf{r}}_{12}\cdot\hat{\mathbf{t}}_{fil}|$ in the top panel. As in Figs.~\ref{global_spin_mass_S} and \ref{alignments_RI}, uncertainties are estimated using bootstrap resampling with 5,000 realizations.
We observe a clear tendency for MMs in the high-mass bin to be preferentially aligned with the filament direction, in agreement with \citet{10.1093/mnrasl/slu106}. We interpret the conversion of orbital angular momentum into the internal spin of the merger remnant as a plausible mechanism driving the pronounced spin reorientations observed in these systems, ultimately leading to spin configurations that are more perpendicular to the filament. To quantify the significance of this signal, we performed a KS test and obtained a p-value of $p=2.09\times10^{-2}$ for the high-mass bin, within the limitations imposed by the small number of halos in this mass range. The bottom panel reproduces the average alignment evolution shown in Fig.~\ref{zmerger}, but considering only halos with a merger-filament alignment greater than 0.7 at the time of the LMM. This selection leaves 182, 28, and 9 halos in the low-, intermediate-, and high-mass bins, respectively. Under this restriction, the trend previously observed only in the high-mass sample becomes apparent also in the intermediate-mass bin, with halos evolving toward more perpendicular spin--filament configurations after the merger event. This result further supports the idea that mergers occurring preferentially along the filament play an important role in driving spin reorientations. Overall, $\sim8\%$ of halos both experience a MM and end in a perpendicular spin--filament configuration.

These results are consistent with previous works \citep[e.g.,][]{10.1111/j.1365-2966.2011.20275.x,10.1093/mnras/stw1395}, which found that MMs produce significant changes in halo spin alignments. Here, we further show that this effect is particularly strong for the most massive halos, where the number of halos is naturally smaller. However, a nuanced interpretation is necessary. While our results confirm that MMs cause the most dramatic individual reorientations (spin swings), they are relatively rare events. As noted by \citet{10.1111/j.1365-2966.2011.20275.x}, approximately 75\% of significant spin changes are driven by the cumulative effect of minor mergers, accretion of small substructures, and flyby encounters, rather than by MMs alone.
Finally, regarding halo shape, MMs drive the system toward a prolate configuration, effectively stretching the mass distribution along the merger axis, consistent with established models \citep{2006MNRAS.367.1781A}.

\section{Discussion}
\label{discussion}

Our results indicate that the evolution of halo spin alignment within filaments is primarily governed by the intrinsic, mass-dependent evolution of halos. In particular, the average spin alignment measured at $z=0$ does not differ significantly from its value at earlier stages ($z=1$), despite the gradual reorientation of specific subpopulations. This behavior is especially evident when halos are separated only by mass or by environment: in both cases, the average alignment curves show that halos tend to preserve their initial orientation. This does not apply to the alignment-evolution subsamples, which are explicitly defined according to their alignment histories.

Major mergers, however, constitute a notable exception to this otherwise smooth evolutionary behavior. We find that mergers generate sharp reorientations of the spin vector and halo shape, accompanied by transient increases in the spin parameter. These features are consistent with the interpretation that mergers inject orbital angular momentum into the remnant halo, which is subsequently redistributed during relaxation. Crucially, as shown by \citet{10.1093/mnrasl/slu106}, mergers preferentially occur along the direction of cosmic filaments. Consequently, the orbital angular momentum of the merging pair is oriented orthogonally to the filament axis. The conversion of this orbital momentum into the internal spin of the remnant drives the notable ``spin swings'' we observe, effectively flipping the halo spin to a perpendicular configuration relative to the cosmic web. We propose that since the most massive halos typically reside closer to the filament core, the deeper local potential well may enhance this coupling, amplifying the dynamical impact of mergers relative to halos in the outskirts.

The physical implications of our results extend naturally to galaxy formation. Observational and theoretical studies suggest that galactic angular momentum partly traces that of the host halo \citep{10.1111/j.1365-2966.2006.11318.x,Tempel_2013}. In this context, our results suggest that the intrinsic, mass-dependent halo evolution identified here provides the gravitational framework for the observed morphological dichotomy, where low-mass spirals align parallel to filaments and massive ellipticals align perpendicularly \citep{10.1093/mnrasl/slu106,KraljicKatarina2021SM3s}. Specifically, while MMs drive the perpendicular reorientation ("spin swings") of massive systems \citep{10.1093/mnrasl/slu106}, smooth accretion is likely to sustain the parallel alignment of low-mass halos by building up angular momentum along the filament direction \citep{Codis_2012,10.1093/mnrasl/slu106}, consistent with the neutral and positive evolutionary tracks identified in this work. Because our analysis is based on DM-only simulations, however, these implications should be interpreted with caution, as baryonic processes can modify the halo--galaxy connection. In particular, hydrodynamical simulations suggest that gas cooling and disk reformation can realign galaxies parallel to filaments, potentially counteracting the merger-induced perpendicular orientation \citep{10.1093/mnrasl/slu106,2019MNRAS.488.4801J}. Quantifying these effects will require a systematic comparison with hydrodynamical simulations.

\section{Conclusions}
\label{conclusion}
In this work, we used zoom-in simulations of ten filaments to study the evolution of the spin--filament alignment of DM halos by tracking their individual and average histories across cosmic time. By separating the alignment curves, we observe that halos tend toward the characteristic orientation of their mass scale. Similarly, the filament environment itself does not introduce significant modifications to the alignment history: halos located inside, near, or outside the filament exhibit comparable evolutionary pathways, with mass consistently remaining the dominant factor.

In contrast, MMs, although relatively rare, produce a clear and temporally localized reorientation of halo spins, also affecting the alignment of the inertia tensor. This suggests that, within the nonlinear regime, the dynamical history of halos -- particularly mergers -- plays a more significant role than the filament itself in determining their spin orientation. Our principal conclusions can be summarized as follows:

\begin{itemize}
\renewcommand\labelitemi{$\bullet$}
    \item Mass is the primary driver of spin--filament alignment. Low-mass halos tend to align parallel to filaments, whereas high-mass halos trend toward perpendicular orientations (though limited by sample size), consistent with the canonical spin-flip behavior \citep{aragon-calvo,Codis_2012}.

    \item Alignment evolution curves reflect intrinsic halo evolution rather than external effects. The different classes -- neutral (25\%), positive (13\%), negative (14\%), and varied (48\%) -- follow mass-driven trends: neutral halos remain preferentially parallel, while negative halos evolve toward perpendicular configurations, consistent with perpendicular alignments arising in the nonlinear regime \citep{2025MNRAS.543.2222L}.

    \item Our environmental characterization, based on the radial distance to the filament spine, reveals no strong dependence of the alignment history on the halo-filament separation. Even halos in the mixed category, which approach the filament during their evolution, do not exhibit systematic changes in their alignment history, suggesting that approaching the filament is not, by itself, associated with significant changes in the spin--filament alignment.

    \item We have investigated MMs as a potential mechanism for reorienting the spin--filament alignment of DM halos. We find that MMs can contribute to reorientations toward perpendicular alignments, but do not appear to be the dominant process, as approximately $17\%$ of halos in the negative evolution class experienced a LMM at $z\leq1$. This suggests that such events are relatively uncommon and that other mechanisms likely dominate the reorientation process. However, when they occur, MMs can be associated with abrupt changes in spin--filament alignment, particularly for higher-mass halos, where the fraction of halos experiencing a MM is about three times higher than in the low-mass bin.

    \item By examining the merger-filament alignment, we find a direct connection between mergers occurring preferentially parallel to the filament and the subsequent reorientation of halo spins toward perpendicular spin--filament configurations. Furthermore, when restricting the sample to mergers with stronger merger-filament alignments, this trend becomes apparent also in intermediate-mass halos.
    
    \item The inertia tensor responds coherently to MMs. As the spin becomes more perpendicular to the filament, the major axis tends to align with the filament while the minor axis evolves toward a more perpendicular configuration, following the spin alignment. This effect is particularly pronounced for the most massive halos \citep{10.1111/j.1365-2966.2011.20275.x,2014MNRAS.441..470T}, which represent the least populated mass regime.

    \item In the nonlinear regime, halo evolutionary history outweighs the large-scale environment. MMs produce the largest individual alignment variations \citep{2002MNRAS.332..325P,10.1093/mnras/stv1570,2015MNRAS.446.2744L}, while the overall misalignment statistics are primarily driven by the cumulative effect of minor interactions and smooth accretion \citep{10.1093/mnras/stw1286,10.1093/mnras/stab411}.

\end{itemize}
We also explored additional dynamical scenarios capable of producing rapid changes in spin alignment, such as crosser halos. Although rare, halos crossing the filament core show transient enhancements in their alignment signal, indicating that filament-crossing events can temporarily modify halo dynamics. A detailed analysis of these systems is beyond the scope of this work and will be addressed in a future study.

Looking ahead, several natural extensions of this work arise. First, our analysis is based on DM-only simulations, and the inclusion of baryonic physics through hydrodynamical simulations will be essential to assess how gas cooling, star formation, and feedback processes modify halo spin acquisition and reorientation, particularly in the highly nonlinear regime explored here. Such effects are expected to be especially relevant for the inner regions of halos, where baryons can significantly reshape the angular momentum distribution \citep{10.1093/mnras/stu1227,10.1093/mnras/stw1286,2020MNRAS.493..362K}.

A closely related aspect is the connection of our results with the framework of IAs, i.e., correlations in galaxy and halo orientations induced by large-scale tidal fields (\citealt{hirata_2004,troxel_2015}; see \citealt{chisari_2025_review} for a complete review). An IAs can bias cosmological parameter inference from weak lensing and clustering measurements and affect cross-correlations with probes such as cosmic microwave background lensing if not properly modeled \citep{hikage_2019,fabbian_2019}. While hydrodynamical simulations show that galaxy spins and shapes exhibit mass- and environment-dependent alignments with the local tidal field \citep{chisari_2015}, our results indicate that halo spin–filament alignment is mainly governed by halo mass and internal dynamical history. In particular, nonlinear events such as MMs produce the strongest localized reorientations, suggesting that IAs originate from early tidal imprints but are subsequently reshaped by nonlinear evolution, including mergers and anisotropic accretion. This highlights the importance of incorporating realistic mass-dependent and nonlinear evolutionary pathways into IA models used to interpret cosmological observations.

Finally, the trends identified here open promising avenues for comparison with upcoming large-scale surveys. Observational programs such as \textit{Euclid} \citep{2025_euclid} and the \textit{Vera C. Rubin} Observatory \citep[LSST;][]{2009arXiv0912.0201L} will provide unprecedented constraints on galaxy orientations, shapes, and their relation to the cosmic web. Establishing a robust connection between halo spin alignments in simulations and observable galaxy properties will be a key step toward testing these theoretical predictions in a cosmological context.

\begin{acknowledgements}
DT acknowledges support from the Programa de Incentivo a la Iniciación Científica (PIIC) from the Universidad Técnica Federico Santa María, grant No.~021/2025. ADMD and DT acknowledge support from the Universidad Técnica Federico Santa María through the Proyecto Interno Regular \texttt{PI\_LIR\_25\_04}. ADMD and DT acknowledge the AstroGainz research group for their support and continuous discussions. RS acknowledges financial support from FONDECYT Regular 2023 project No. 1230441 and also gratefully acknowledges financial support from ANID - MILENIO NCN2024\_112.
\end{acknowledgements}

\bibliographystyle{aa}
\bibliography{references}

\end{document}